\documentclass[10pt,conference]{IEEEtran}
\IEEEoverridecommandlockouts
\usepackage{cite}
\usepackage{amsmath,amssymb,amsfonts}
\usepackage{algorithmic}
\usepackage{graphicx}
\usepackage{svg}
\usepackage{lipsum} 
\usepackage{textcomp}

\usepackage[english]{babel}
\usepackage{lipsum}
\usepackage{soul}
\usepackage{multirow}
\usepackage{tikz}
\usepackage{mdframed}
\usepackage{pgfplots}
\usepackage{soul}

\definecolor{mygray}{RGB}{211,211,211} %

\makeatletter
\newcommand{\linebreakand}{%
  \end{@IEEEauthorhalign}
  \hfill\mbox{}\par
  \mbox{}\hfill\begin{@IEEEauthorhalign}
}
\makeatother

\def\BibTeX{{\rm B\kern-.05em{\sc i\kern-.025em b}\kern-.08em
    T\kern-.1667em\lower.7ex\hbox{E}\kern-.125emX}}
\begin{document}

\title{Towards Emotionally Intelligent Software Engineers: Understanding Students' Self-Perceptions After a Cooperative Learning Experience
\thanks{This study was financed in part by the Coordenação de Aperfeiçoamento de Nível Superior - Brasil (CAPES) - Postdoctoral Institutional Program (PIPD) and the Conselho Nacional de Desenvolvimento Científico e Tecnológico - Brasil (CNPq) - Universal grant 404406/2023-8.}
}

\author{\IEEEauthorblockN{1\textsuperscript{st} Allysson Allex Araújo}
\IEEEauthorblockA{\textit{Center for Science and Technology} \\
\textit{Federal University of Cariri}\\
Juazeiro do Norte, Brazil \\
allysson.araujo@ufca.edu.br}
\and
\IEEEauthorblockN{2\textsuperscript{nd} Marcos Kalinowski}
\IEEEauthorblockA{\textit{Department of Informatics} \\
\textit{Pontifical Catholic University of Rio de Janeiro}\\
Rio de Janeiro, Brazil \\
kalinowski@inf.puc-rio.br}
\and
\IEEEauthorblockN{3\textsuperscript{rd} Matheus Paixao}
\IEEEauthorblockA{\textit{Graduate Program in Computer Science} \\
\textit{State University of Ceará}\\
Fortaleza, Brazil \\
matheus.paixao@uece.br}
\and
\linebreakand
\IEEEauthorblockN{4\textsuperscript{th} Daniel Graziotin}
\IEEEauthorblockA{\textit{Chair of Information Systems and Digital Technologies} \\
\textit{University of Hohenheim}\\
Stuttgart, Germany \\
graziotin@uni-hohenheim.de}
}

\maketitle

\begin{abstract}
[Background] Emotional Intelligence (EI) can impact Software Engineering (SE) outcomes through improved team communication, conflict resolution, and stress management. SE workers face increasing pressure to develop both technical and interpersonal skills, as modern software development emphasizes collaborative work and complex team interactions. Despite EI's documented importance in professional practice, SE education continues to prioritize technical knowledge over emotional and social competencies.
[Objective] This paper analyzes SE students' self-perceptions of their EI after a two-month cooperative learning project, using Mayer and Salovey's four-ability model to examine how students handle emotions in collaborative development.
[Method] We conducted a case study with 29 SE students organized into four squads within a project-based learning course, collecting data through questionnaires and focus groups that included brainwriting and sharing circles, then analyzing the data using descriptive statistics and open coding.
[Results] Students demonstrated stronger abilities in managing their own emotions compared to interpreting others' emotional states. Despite limited formal EI training, they developed informal strategies for emotional management, including structured planning and peer support networks, which they connected to improved productivity and conflict resolution.
[Conclusion] This study shows how SE students perceive EI in a collaborative learning context and provides evidence-based insights into the important role of emotional competencies in SE education.
\end{abstract}


\section{Introduction} %

Emotional Intelligence (EI) is a fundamental aspect of human interaction and personal development, influencing our ability to manage social complexities and adapt to changing environments \cite{mayer2008human}. EI encompasses self-awareness of one's own goals, intentions, and behaviors, as well as the understanding of others' emotions and actions, significantly affecting our behavior and relationships \cite{mayer2001emotional}. Introduced in 1990 by Salovey and Mayer \cite{salovey1990emotional}, EI gained widespread attention through Daniel Goleman's influential book \cite{goleman1996emotional}. As MacCann et al. \cite{maccann2020emotional} explain, EI extends beyond basic emotional awareness, representing a complex interplay between personal and social capabilities. This concept has changed our understanding of human potential, challenging traditional views of intelligence and success.

EI's influence extends beyond personal relationships, notably impacting professional success. Recent meta-analyses show that individuals with high EI consistently perform better than their peers across various job roles \cite{joseph2015does}. High EI improves communication and conflict resolution skills \cite{jorfi2014impact}, promotes empathy \cite{ioannidou2008empathy}, and often results in more effective leadership \cite{sadri2012emotional}. These EI-related benefits are particularly important in Software Engineering (SE), where successful system development depends on effective teamwork and constructive interactions among diverse groups of practitioners \cite{novielli2019sentiment}.

The importance of an emotion-focused perspective has gained attention among SE researchers in recent years \cite{wrobel2013emotions, fountaine2017emotional, novielli2023emotion}. Studies have examined the connection between emotions and developers' productivity \cite{girardi2021emotions}, emotional triggers in software development \cite{girardi2020recognizing}, EI in SE education \cite{hidalgo2004use}, and the impact of emotions on developers' well-being \cite{graziotin2018happens}. Research has also investigated EI's role in managing requirements changes \cite{madampe2022role} and its relation to work preferences in SE \cite{kosti2014personality}.

Considering the importance of EI and emotions in SE, an important question arises: How well are we preparing future software engineers in terms of EI? Understanding SE students' perspectives on EI can identify gaps between industry-valued skills and academic development, provide insights into students' preparedness for complex team dynamics, and inform the development of focused EI training programs. 

While EI skills can be developed on the job, university years offer a prime opportunity to establish a strong EI foundation alongside technical skills in a learning-focused environment. Early focus on EI development can produce long-term benefits~\cite{mattingly2019can}. Including EI training in SE curricula can prepare students for professional challenges, potentially improving job satisfaction and performance \cite{joseph2015does}. Furthermore, the fast-paced and high-pressure software industry environment may not always support systematic EI skill development. Recent research indicates that software engineers face work-related stress and interpersonal conflicts \cite{graziotin2018happens,ostberg2020methodology}, worsened by overwhelming settings \cite{michels2024}. These issues could be mitigated with better EI skills. As SE moves towards more collaborative and globally distributed work models \cite{DeSouza2023}, the need for professionals with strong EI skills is expected to grow.

Including EI development in SE education could help to approach current industry needs and better prepare students for the future of software development. We address the following research question: \textbf{\textit{What are the self-perceptions of Software Engineering students regarding their Emotional Intelligence after participating in a cooperative learning experience?}} To answer the question, we investigate SE students' self-perceptions of EI after a two-month cooperative learning experience. This evaluative case study was organized in the form of a project-based learning course, where 29 students (grouped into four squads) cooperatively learned subjects such as User Experience, React, or Spring while working to solve a real-world problem. This collaborative environment, supported by a defined learning objective, sociotechnical interactions, and expectations, served as a valuable case study to examine students' perceptions on their EI. To gather and analyze data, we administered a questionnaire followed by four online focus group sessions conducted at the end of the course. These sessions also incorporated brainwriting and sharing circle methods. The data was then analyzed quantitatively using descriptive statistics and qualitatively through open coding.

Our study contributes both to research and practice. Academically, we analyze through the lenses of an EI framework\cite{mayer1997emotional}, how SE students manage emotions during cooperative learning, exploring their emotional regulation strategies and their impact on performance. For industry, we offer insights into developing EI training programs for SE novices, aiming to enhance the emotional preparedness of the future workforce.

The paper is structured as follows: Section 2 reviews background and related work; Section 3 details the case study design, followed by a presentation of results (Section 4) and discussion (Section 5). The paper concludes in Section 6.

\section{Background and Related work}
\subsubsection{EI and its Relevance for SE Education and Practice} EI encompasses self-awareness of one’s own goals, intentions, and behaviors, as well as the understanding of others' emotions and actions \cite{mayer2008human}. Mayer and Salovey \cite{mayer1997emotional} posited that EI comprises four abilities: (1) the ability to perceive emotions in oneself and others, as well as in objects, art, and stories (perception of emotion), (2) the ability to generate emotions in order to use them in other mental processes (emotional facilitation of thought), (3) the ability to understand and reason about emotional information and how emotions combine and progress through relationship transitions (understanding emotions), and (4) the ability to be open to emotions and to moderate them in oneself and others (managing emotions). %

EI has gained increasing relevance in SE, where its impact on developers' productivity, emotional triggers, and well-being is a growing area of interest \cite{novielli2019sentiment}. Two decades ago, Hidalgo et al. \cite{hidalgo2004use} made early efforts to integrate EI into SE education by embedding EI skills into a course structure, though their work primarily focused on course design rather than student perspectives. More recently, Arroyo-Herrera et al. \cite{arroyo2019development} launched the "Experiencia 360º" project, employing action research and gaming techniques to strengthen socio-emotional skills among Computing Engineering students.

\subsubsection{Emotional Awareness in SE} 
Kosti et al. \cite{kosti2014personality} examined personality clusters among software engineers, revealing how distinct personality traits correlate with work preferences. Building on the intersection of EI and work performance, Khatun and Salleh \cite{Khatun2020moderation} employed hierarchical multiple regression analysis to investigate the moderating role of EI in the relationship between work ethics and performance among software engineers. Wrobel \cite{wrobel2013emotions} surveyed 49 software developers, revealing how emotions affect performance and introducing ``emotional risk" to productivity.

More recently, Novielli and Serebrenik \cite{novielli2023emotion} assessed emotions in software ecosystems, revealing how developer emotions can indicate ecosystem health and development issues. This review also summarized tools and datasets for emotion analysis. In another study, Serebrenik \cite{serebrenik2017emotional} shifted the focus to the emotional labor experienced by software developers. Despite stereotypes of developers as isolated workers, the study argued that the growing social demands in software development make emotional labor increasingly relevant. 

The role of emotional awareness in software development and productivity was further explored by Fountaine and Sharif\cite{fountaine2017emotional}, who investigated how emotional clarity affects task progress. Their study proposed that enhancing emotional awareness in developer environments could improve productivity. In addition, Girardi et al. \cite{girardi2020recognizing} conducted a study with 23 students to identify emotional triggers and coping strategies during programming tasks. Their research highlighted the effectiveness of wristbands measuring electrodermal activity and heart metrics in recognizing emotions, contributing to the development of emotion recognition classifiers. In a follow-up study, Girardi et al. \cite{girardi2021emotions} examined the impact of emotions on perceived productivity among 21 professionals from five companies. Replicating Graziotin et al.~\cite{graziotin2015}, they confirmed a positive correlation between emotional valence and productivity, especially in the afternoon, highlighting the importance of individualized emotion detection models.

Graziotin et al. \cite{graziotin2018happens} explored the impact of happiness and unhappiness on software developers during the development process. Analyzing responses from 317 participants, they identified 42 consequences of unhappiness and 32 of happiness, influencing cognitive and behavioral states as well as external outcomes. Also related to the perception and management of one's emotions, Guenes et al. \cite{guenes2024impostor} examined the impostor phenomenon in software engineers. From responses of 624 participants, they found that frequent and intense impostor feelings led to lower self-perceived productivity.

\subsubsection{Emotions in Agile and Socio-Technical Settings} Luong et al. \cite{luong2021agile} investigated EI in addressing human-related challenges within Agile Managed Information Systems projects. Analyzing data from 194 agile practitioners, they found significant correlations between EI and key issues such as anxiety, motivation, trust, and communication within agile teams. This intersection of EI in agile contexts, particularly focusing on requirements change handling, was further expanded by Madampe et al. \cite{madampe2022emotional}, who studied how software practitioners emotionally react to requirements changes throughout their lifecycle, based on a global survey of 201 practitioners. They identified common emotional responses and triggers,  highlighting the negative impact of last-minute changes near deadlines. Madampe et al. \cite{madampe2023framework} developed a framework for handling emotion-oriented requirements changes, based on a survey of 241 participants, identifying emotional challenges and mitigation strategies in socio-technical environments. Their later work linked negative emotions in agile projects to reduced developer satisfaction and productivity, offering practical solutions. Additionally, they showed that EI and cognitive intelligence are key to managing requirements changes in agile settings, using a mixed-method study with 124 practitioners to suggest strategies for emotion regulation, relationship management, and maintaining productivity \cite{madampe2024supporting}.

\subsubsection{Research Gap} 
In summary, early initiatives and recent studies in SE highlight the relevant role of EI in developers' well-being and productivity. However, most research emphasizes external observations, leaving a gap in understanding SE students' self-perceptions of EI. While professional settings have been explored, little attention has been given to how EI is developed and perceived in educational contexts. Specifically, research on how SE students evaluate their own EI through Mayer and Salovey's framework \cite{mayer1997emotional} remains limited. This study aims to address these gaps by investigating SE students' self-perceptions of EI within a project-based learning context.

\section{Method}

\subsubsection{Research Design}
\label{subsubsec:context}

This evaluative case study adheres to the guidelines outlined by Runeson and Höst \cite{runeson2009guidelines}, which emphasize five key aspects: research question (stated in the Introduction), case study selection (including unit of analysis), data collection, data analysis, and validity evaluation. %

This investigation concerns a case study with SE students as the \textbf{unit of analysis} for assessing their EI based on the abilities defined by Mayer and Salovey \cite{mayer1997emotional}. The \textbf{case selection} involved a project-based learning course centered on developing a web platform to support tourism in the inland regions of Ceará state, in Brazil. This case was purposefully selected due to its alignment with the research aim of investigating SE students' self-perceptions of EI after participating in a cooperative learning experience. The project-based nature of the course provided an ideal context for analyzing emotional and collaborative dynamics. The participants were selected from students interested in enrolling in the course, and all of them signed an informed consent form.

\textbf{Data collection} involved a questionnaire and conducting four virtual focus group sessions, one for each squad, after the course. The questionnaire collected demographic and academic information (age, course, enrollment duration) and addressed key questions: previous experiences with cooperative learning and EI training, familiarity with EI (rated 1 to 5), self-assessed ability to identify and express emotions in teamwork (rated 1 to 5), stress management strategies (multiple-choice scale with options ranging from feeling easily overwhelmed and unmotivated to being unaffected by stress and remaining focused and productive), belief in EI’s impact on academic performance and teamwork (rated 1 to 5), a specific situation where EI was important for success in cooperative learning, and strategies for managing conflict and emotional challenges in teamwork. EI self-assessment items were inspired from Mayer and Salovey’s model \cite{mayer1997emotional}.

Regarding the focus groups sessions, we adopted the brainwriting and sharing circle methods to enhance the discussions\footnote{Brainwriting is a brainstorming technique where participants generate and share written ideas in a timed sequence \cite{heslin2009better}. The sharing circle is a roundtable discussion where participants take turns sharing thoughts on a topic \cite{tachine2016sharing}.}. Each focus group was conducted separately for each of the four squads, and all sessions were held online and moderated by the first author. We chose virtual focus groups because they enable the collection of diverse perspectives in a setting that allows for immediate feedback and interaction among participants \cite{krueger2014focus}. In line with the best practices for focus groups \cite{morgan1996focus}, the moderator aimed to provoke participants to explore the factors underpinning their responses. 

The focus group sessions were recorded and structured in three steps: 1) A brief introduction to EI using a slide deck, covering the basics of EI, its importance, and ways to develop it; 2) The discussion was structured around the four abilities defined by Mayer and Salovey \cite{mayer1997emotional}. Using the Miro tool, we created four columns with four questions, each representing one of the abilities. The students freely choose rows to write their answers in the form of sticky notes. Thus, before discussing one ability, students wrote their thoughts on a virtual sticky note in the appropriate column. After everyone had contributed, the group engaged in a sharing circle to collectively discuss their responses, fostering a reflective environment. This process was repeated for all four abilities; 3) The session concluded with a summary of the main points, an invitation for final comments, and thanks to the participants. %

\textbf{Data analysis} involved descriptive statistics for quantitative items (closed-ended questionnaire responses) and open coding for qualitative data (focus groups and open-ended questionnaire responses). For open-ended questionnaire data, we used an inductive coding approach to allow themes to emerge naturally, while focus group data was analyzed deductively, guided by Mayer and Salovey's EI framework \cite{mayer1997emotional}. One co-author conducted the initial coding independently, and two additional co-authors iteratively reviewed and refined the set of codes, resolving discrepancies through discussion. This collaborative process ensured rigor and triangulation, enabling us to identify recurring themes in students' self-perceptions of EI and interpret them in relation to our research questions.

To strengthen the \textbf{validity of our findings}, we adhered to established case study best practices \cite{runeson2009guidelines, ralph2020empirical}. Specifically, we employed data triangulation by integrating theoretical insights (primarily drawn from Mayer and Salovey \cite{mayer1997emotional}) with quantitative and qualitative results obtained from two distinct data collection methods: questionnaires and focus groups. The focus groups, in particular, utilized techniques such as brainwriting and sharing circles to enrich the discussions. This interpretation based on descriptive statistics and open coding allowed cross-validation of findings. Finally, all data supporting this study (codebook, questionnaire answers, etc.) is openly accessible through our repository \cite{repo}. 

\subsubsection{Context and Participants' Description}
\label{subsubsec:context}
The course was structured around the principles of project-based learning, flipped classroom, cooperative learning, and gamification (using Kahoot quizzes). While the teaching itself did not explicitly focus on EI, aspects such as teamwork, feedback, and communication were inherently part of the project development. Students were grouped into four squads based on their study interests, previously informed to the professor: User Experience (UX), React, or Spring. Each squad followed a tailored curriculum with curated content suggested by the professor, who also served as the course coordinator and Product Owner. The project centered on developing a web platform to promote regional development through tourism in the Brazil's inland. The course spanned two months, with weekly in-person or online meetings. For each session, students were assigned content to study beforehand (as suggested by flipped classroom), and one student acted as the facilitator. The facilitator was responsible for leading the discussion, creating a Kahoot quiz, and preparing an implementation challenge. Additionally, the squads had to cooperatively plan and discuss the weekly development tasks for the ongoing project.

Regarding the characterization of the participants, we gathered information from 25 students with diverse academic backgrounds: 52\% studied computer science, and 32\% studied information systems. Additionally, three participants (12\%) were high school students enrolled in an informatics technical program and not yet university students. Participants' ages ranged from 17 to 37 years, with the majority (64\%) aged between 17 and 22. Gender distribution showed a predominance of male participants, with 60\% identifying as male and 40\% as female. Four (out of 29) students did not respond to the questionnaire, and seven did not attend the focus group sessions due to schedule conflicts.

The duration of university enrollment varied, with 28\% having been enrolled for less than a year, 48\% for one to three years, and 12\% for more than three years. The remaining 12\% were not enrolled in the university since they were high school students. A large portion (80\%) had no prior experience with cooperative learning activities before the course, while 20\% had some previous engagement, mostly informal and minimally structured, typically involving basic peer collaboration with colleagues. In terms of EI training, 80\% of participants had not undergone any prior training, whereas 20\% had some exposure, such as high school workshops, mentorship programs, or online courses.

\section{Results and Analysis}

\subsection{Questionnaire Analysis}
\label{subsubsec:survey}

\begin{figure}[ht!]
	\centering
	\includegraphics[scale=0.52]{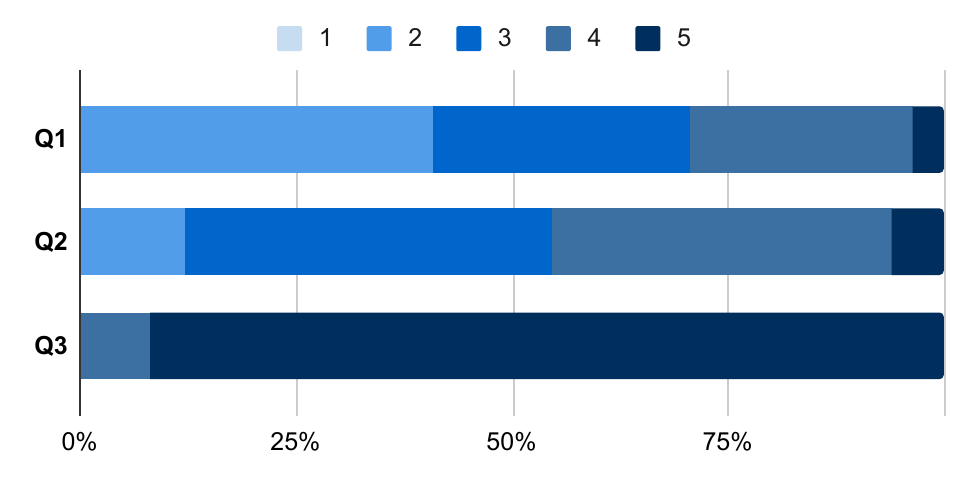}
	\caption{Familiarity with EI (Q1), self-assessment of EI (Q2), and views on EI’s impact on academic performance and teamwork  (Q3).}
	\label{fig:chart1}
\end{figure}

Figure \ref{fig:chart1} provides an overview of the students' responses to three closed-ended questions, respectively covering their familiarity with EI (Q1), their self-assessment of emotional skills (Q2), and their views on EI's impact on academic performance and teamwork in software development (Q3). Participants evaluated their familiarity with EI on a scale from 1 to 5, where 1 denoted no knowledge and 5 signified expert-level understanding. The scores ranged from 2 to 5, with a mean of 3.30 (SD=.72), indicating a moderate grasp of EI concepts. This result suggests that while most participants had some awareness of EI, there is room for improvement. When assessing their own abilities to identify and express emotions within team settings, the scores also ranged from 2 to 5, with a mean of 3.40 (SD=.51), reflecting a relatively high self-perceived competency in emotional awareness and expression, although individual scores vary. Finally, participants evaluated the impact of EI on academic performance and teamwork, with ratings ranging from 4 to 5 and a mean of 4.88 (SD=.33). This consensus highlights a strong belief in the positive influence of EI on both academic success and effective teamwork.

\begin{figure}[ht!]
	\centering
	\includegraphics[width=\columnwidth]{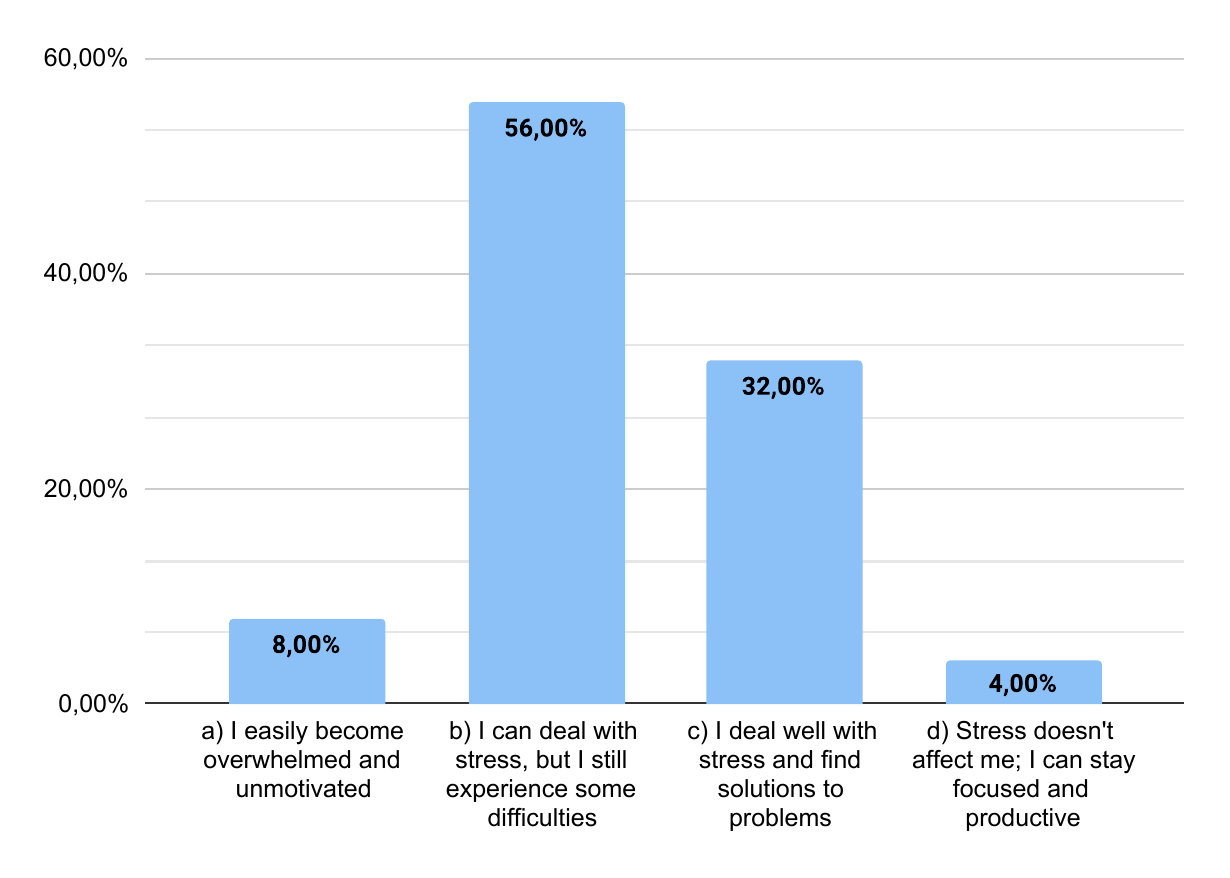}
	\caption{How the students manage stress and pressure within a cooperative learning environment?}
	\label{fig:chart2}
\end{figure}

Students were asked on how they managed stress and pressure within the cooperative learning environment. As shown in Figure \ref{fig:chart2}, 56\% (14 participants) reported that they can manage stress, albeit with challenges. Additionally, 32\% (8 participants) indicated that they handle stress effectively and are able to find solutions to problems. In contrast, 8\% (2 participants) felt that they become easily overwhelmed and demotivated by stress, while only 4\% (1 participant) stated that stress does not impact their focus or productivity. %

\begin{figure*}[ht!]
	\centering
	\includegraphics[scale=0.17]{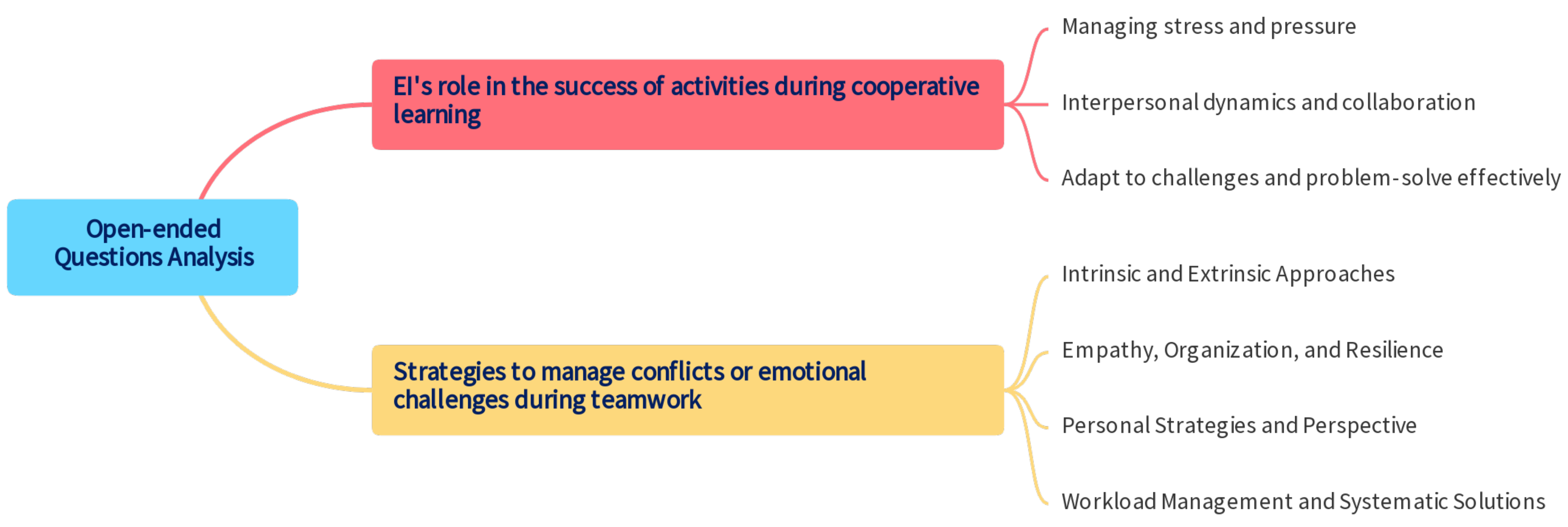}
	\caption{Open coding of situations in which i) EI played a important role in the success of activities during cooperative learning (first theme) and ii) strategies students employ to manage conflicts or emotional challenges that arise during teamwork (second theme).}
	\label{fig:chart3}
\end{figure*}

Regarding the open-ended questionnaire responses, the \textbf{codes} derived from the analysis are summarized in Figure \ref{fig:chart3}, categorized into two main themes: EI's role in the success of activities during cooperative learning and Strategies for addressing conflicts or emotional challenges in teamwork. Within the first theme, students were asked to describe specific situations where EI played an important role in the success of activities. Participants frequently highlighted how EI was relevant for \textbf{managing stress and pressure}. For instance, P2 recounted, ``On the day I was a facilitator, I had to concentrate well to present because I am an anxious person''. Similarly, P12 shared, ``I was very nervous during my turn as the facilitator; I wanted to give up, but I took deep breaths and reminded myself that we all attended the same classes and were there to share our learning''. P20 highlighted how having supportive colleagues made difference: ``When I was overwhelmed with some tasks, it was really helpful to have colleagues assisting and showing me how to respond, because when I'm overwhelmed, I end up not being very productive. Having a way to handle that, and especially having people supporting me, makes a big difference''.

According to the students, \textbf{interpersonal dynamics and collaboration} emerged as important issues where EI made an impact. P7 described handling a challenging situation where a team member tried to assert dominance uncomfortably: ``When someone wanted to show off their knowledge in an uncomfortable way, we had to intervene without making them feel embarrassed, as everyone was there to learn equally''. P15 provided an example of supporting a nervous colleague: ``One time, a colleague who was the pilot was visibly nervous during his turn to deliver the content. He had a deadline for another assignment close to the time of our session and maybe hadn’t been able to fully grasp the material to present it. We helped him complete the lesson by participating, making contributions, and he managed to finish it''. P19 weighted on the importance of emotional support from peers: ``Recently, I went through a difficult time, and if it hadn’t been for some of my colleagues, I would have lost focus''.

The ability to \textbf{adapt to challenges and solve problems effectively} also highlighted the importance of EI. For example, P10 recounted a situation involving a technical issue: ``At one point, the CSS code got broken, but instead of losing my cool, I gathered everyone to talk and figure out who changed what and how we could fix it together''. P13 talked about understanding team dynamics : ``It was the first time I worked on a design prototype as a team, and I realized that emotional intelligence is important for managing different ideas and decisions''. P18 also highlighted the challenge of balancing commitments: ``Balancing university, internship, personal studies, and the  meetings required effective emotional management''. P4 mentioned ``in terms of deadlines and expectations''. P6 reflected on the end-of-semester pressures: ``I believe emotional intelligence was important during the exam and project period at the end of the semester, as I had to handle many things in a short time while still trying to be present at the team meetings''. P24 concluded, ``[...] emotional intelligence is very important at work. Mixing personal problems with work negatively affects productivity''.

The second theme covered the students' strategies to manage conflicts or emotional challenges during teamwork. The \textbf{codes} revealed the following solutions:  

\begin{itemize}
\item \textbf{Intrinsic and Extrinsic Approaches}: P2 mentioned mental preparation: ``Before entering the work zone, I try to clear my mind of problems''. P3 emphasized: ``Dialogue''. P4 turned stress into motivation, noting, ``I use stress as motivation to achieve my goals''. P5 combined breathing and communication, and P6 sought opinions from others to better express himself.

\item \textbf{Empathy, Organization, and Resilience}: P7, P8, P9, and P16 highlighted empathy and organizational techniques. P7 used empathy and corrective feedback, ``Bringing empathy into play or pointing out necessary areas where mistakes are being made without naming individuals''. P8 divided tasks to reduce tension, and P9 took breaks to manage stress: "[...] I set it aside, and once I'm feeling better, I come back to it''.  P16 emphasized resilience: ``I try to focus on the present, which helps me organize my thoughts and concentrate better on what I need to do''.

\item \textbf{Personal Strategies and Perspective}: P10 explained: `` I'm very calm when it comes to working in a team; even though I may not talk much or show it, if I see a problem, I try to speak with someone who can communicate it better than I can to the whole team''. P11 used music and focused on task completion, ``I like to listen to music when I'm alone and think about how I can help''.  P12 found brief solitude helpful, ``I spend a short period of time alone to organize my thoughts, calm down, and think about possible solutions''. P19 stressed empathy, ``Always listen to others' perspectives and put yourself in their shoes'', while P21 focused on prioritization: ``Focus on what is important and try to abstract away the rest''.

\item \textbf{Workload Management and Systematic Solutions}: P23 created schedules to manage tasks, ``I try to create a schedule to organize myself better because stress usually comes from having too many things to do in a short period of time''. P24 gained perspective by viewing situations impartially, ``I try to calm myself and view the situation from above''. P25 motivated self by problem-solving, ``I try to self-motivate by thinking of ideas that can help solve the problems''.
\end{itemize}

\subsection{Focus Groups Analysis}
\label{subsubsec:fg}

We conducted four virtual focus groups, each representing one of the four squads involved in the study. The focus group with first UX Squad, consisting of 3 participants, was held on 01/08/2023 and lasted 125 minutes. The focus group with the second UX Squad\footnote{There were two UX squads due to higher student interest in this area.}, comprising 7 participants, took place on 31/07/2023 for 150 minutes. The focus groups with the React Squad, which included 4 participants, met on 19/08/2023 for 105 minutes. Finally, the focus group with the Spring Squad, consisting of 8 participants, convened on 26/08/2023 for 120 minutes. On average, the sessions lasted 125 minutes. These sessions, which explored methods such as brainwriting and sharing circles, addressed four main perspectives aligned with the EI abilities proposed by Mayer and Salovey \cite{mayer1997emotional}:

\begin{itemize}
    \item \textit{Perceiving Emotions}: During teamwork activities, how did you identify and express your own emotions and those of your colleagues? Was there any situation where the perception of emotions positively influenced collaboration and decision-making? \\ 
    \textit{Rationale}: By exploring how students perceive emotions in the context of SE, we can gain insights into their ability to create an emotionally intelligent environment;
    \item \textit{Using Emotions}: How did you use positive emotions to enhance creativity and problem-solving? In what ways can incorporating emotions into the thinking process lead to higher-quality solutions? \\
    \textit{Rationale}: By understanding how students access and utilize positive emotions to drive creativity and innovation, we can better comprehend how EI contributes to the development of more effective and creative skills;
    \item \textit{Understanding Emotions}: How did you interpret and understand your colleagues' emotions when technical challenges arose during teamwork? How did this emotional understanding influence collaboration and the pursuit of joint solutions? \\
    \textit{Rationale}: By understanding how students interpret and respond to their colleagues' emotions, we can identify the importance of EI in creating a healthy and productive team environment;
    \item \textit{Managing Emotions}: How did you deal with frustration and stress during tight deadlines and intense demands? What strategies did you use to regulate your emotions and maintain focus on progress and work quality? \\
    \textit{Rationale}: By understanding how students cope with these emotions and regulate their reactions, we can gain insights into the development of emotional management.
\end{itemize}

\begin{figure*}[ht!]
	\centering
 \includegraphics[scale=0.16]{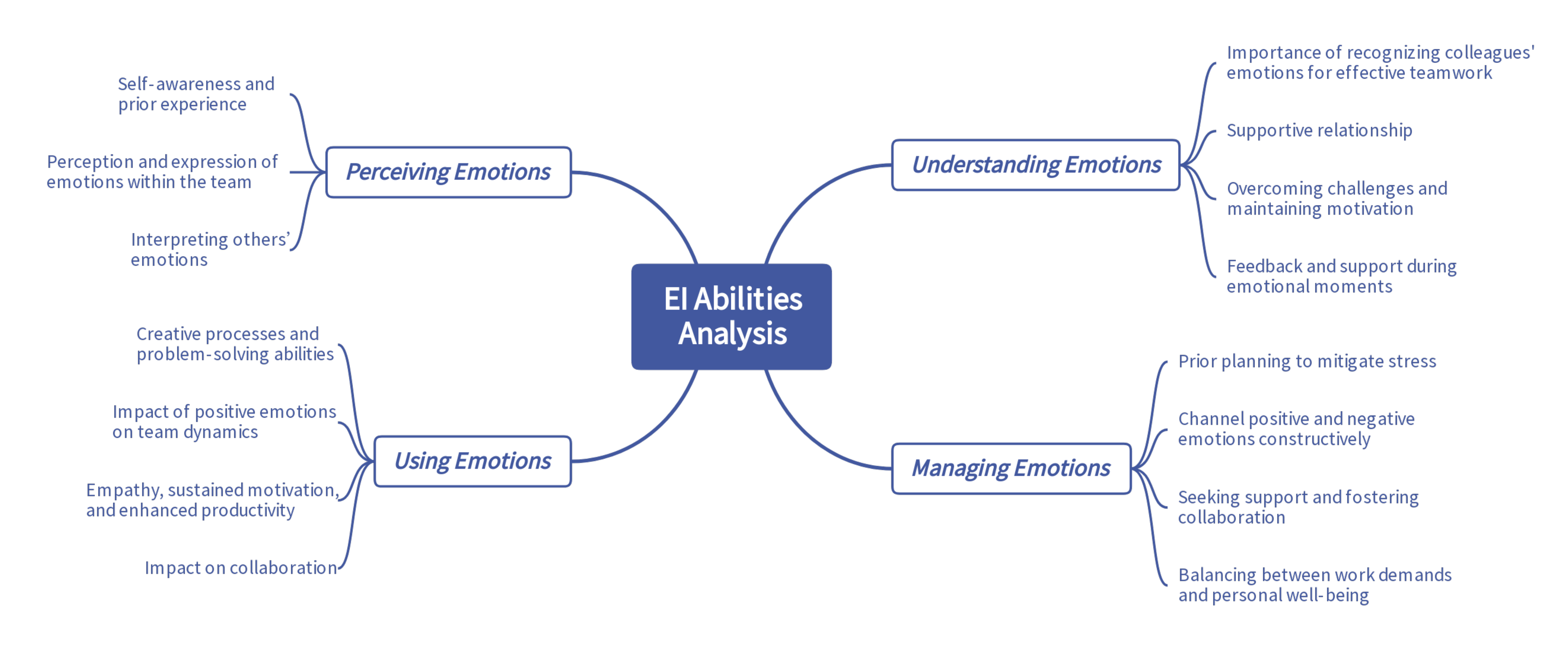}
	\caption{Open coding about the four seminal EI abilities identified by Mayer and Salovey \cite{mayer1997emotional}.}
	\label{fig:chart4}
\end{figure*}

Figure \ref{fig:chart4} displays the \textbf{codes} derived from our analysis of the EI abilities (\textit{themes}) proposed by Mayer and Salovey \cite{mayer1997emotional}. We discuss each ability from the students' perspective below.

\subsubsection{Perceiving Emotions (the ability to perceive accurately, appraise and express emotion)}

Students found it straightforward to identify their own emotions, often attributing this to \textbf{self-awareness and prior experience}. For example, P1 noted, ``My emotions were easy to identify, and the emotions from the team positively influenced some parts of the project''. P2 echoed this sentiment: ``My own emotions were easier to identify because of self-awareness, but understanding my colleagues’ emotions, especially those I didn’t know beforehand, was more challenging''. While self-awareness aids in emotional perception, recognizing the emotions of less familiar team members may require additional effort.

Regarding the \textbf{perception and expression of emotions within the team}, participants noted an impact on the project. P3 remarked, ``My emotions were expressed through how I communicated, and I identified my colleagues' emotions through dialogue and observation''. P6 observed, ``My emotions were easy to identify and express. In the prototyping phase, expressing emotions was a strong point, as everyone seemed to feel comfortable expressing their opinions''. These issues demonstrate how recognizing and addressing emotional states can influence project outcomes.

In particular, \textbf{interpreting others' emotions} also emerged as another challenge. P7 shared, ``Identifying others' emotions was even harder. Sometimes, it wasn't about words but rather looks or gestures that indicated whether things were going well''. This issue highlights the importance of non-verbal cues in emotional perception. The ability to perceive emotions has also impacted team interactions. P14 recalled, ``In one meeting, we had to support a colleague to stay calm so we could proceed. It was a team effort to identify and support him with his emotions''. Similarly, P17 noted, ``Countless times during the development of the project, it was necessary to have conversations beyond the scope, like discussions about the future and value, which greatly helped with the project’s productivity''. These examples illustrate how EI may contribute to long-term team effectiveness.

\subsubsection{Using Emotions (the ability to access and/or generate feelings when they facilitate thought)}

Participants mentioned that positive emotions contributed to their \textbf{creative processes and problem-solving abilities}. P1 observed, ``[positive emotions] help to see the problem from a different perspective, and even find something positive in a challenge. Emotions have a strong influence in any aspect, and the better we manage them, the better we can overcome obstacles''. P4 similarly highlighted how motivation, driven by positive emotions, spurred diverse perspectives: ``I took advantage of moments of motivation to seek different perspectives and inspiration, as well as focus on the parts of the project that demanded more from me''. These perceptions indicate that leveraging positive emotions can lead to more comprehensive and creative problem-solving by enhancing cognitive flexibility.

We also found the \textbf{impact of positive emotions on team dynamics}. In this sense, P5 noted, ``Positive emotions helped with communication and the project development process. A good example is when colleagues complemented each other's statements regarding the content''. In addition, P10 reflected, ``Positive emotions, especially those that emerged at the beginning of the course, helped fuel the desire to deliver something well-done, creative, and presentable during the classes as a facilitator. Harnessing these positive emotions aids in making good decisions and boosting creativity''. This finding emphasizes the synergy between positive emotional states and effective decision-making, suggesting that emotional positivity can enhance the quality of work.

Our analysis revealed the role of \textbf{empathy, sustained motivation, and enhanced productivity} in achieving successful outcomes. P13 emphasized: ``The main method was to listen to everyone’s ideas, which allowed us to come up with various creative solutions, leading us to a great final result. Incorporating emotions into the thought process brings about a more empathetic solution''. Additionally, P16 noted the importance of a positive emotional state for motivation and creativity: ``By seeking to extract the best from them, especially from the positive emotions that generate a lot of motivation and help maintain enthusiasm for the project, creativity is also enhanced, leading to new ideas and solutions''. 

Furthermore, positive emotions were observed to have a \textbf{impact on collaboration}. P17 observed, ``When there is a pleasant environment and good interaction, it becomes much easier to develop and produce higher-quality code. During the project, we often helped each other, which made the development process easier and much more productive''. This code illustrates how a positive emotional atmosphere can enhance collaboration. P21 added, ``Transforming these positive emotions into focus and concentration to perform well. Once you effectively incorporate emotions, you gain more control over your brain, which helps you arrive at more accurate solutions''. This finding suggests that managing positive emotions may improve cognitive control and problem-solving.

\subsubsection{Understanding Emotions (the ability to understand emotion and emotional knowledge)}

Students emphasized the importance of \textbf{recognizing colleagues' emotions for effective teamwork}. P2 highlighted the relevance of perceiving emotions, especially during stressful periods: ``It was very important to recognize the emotions of colleagues, especially at the end of the semester when most were overwhelmed with exams and final assignments''. Similarly, P3 described how observing emotions through expressions and dialogue helped understand the colleagues: ``Emotions were interpreted through observation and dialogue, as it was possible to identify whether someone was feeling good about a particular task based on their expressions. Emotional awareness was very helpful in dealing with people because it allowed us to understand what they enjoyed doing and what made them uncomfortable''.

This emotional awareness fostered a more \textbf{supportive relationship} among the students alongside the course and project. P4 stressed the importance of empathy towards each team member's schedule and limitations: ``We made a point to understand each other's schedules and constraints. This empathy was essential in ensuring that solutions were workable for everyone.'' P5 added that addressing colleagues' stress strengthened relationships and commitment: ``It was interpreted in a way that prevented the person from becoming more stressed about it, as this helped not only with commitment but also in strengthening relationships''. In addition, P14 related that friendships formed during the project facilitated better communication and support: ``As we developed friendships, it became easier to support each other when issues arose. This understanding made our interactions more dynamic''.

Another code involved the role of emotions in \textbf{overcoming challenges and maintaining motivation}. In this context, P21 shared that  emotions from colleagues fueled motivation: ``I noticed in everyone a strong desire to overcome challenges, reflected in their focus and collaboration. Seeing the team's motivation ends up inspiring us to follow the same path with even more determination''. P18 illustrated how understanding a colleague's motivation was key to providing effective support: ``[...] a colleague needed help, and since I already had some knowledge of the technology, I was able to suggest what he could look into to solve his challenge''.

Furthermore, providing \textbf{feedback and support during emotional moments} was another important issue for improving team morale. P19 mentioned using positive feedback to calm colleagues before presentations: ``I think the moment when people show their emotions the most is during presentations, where they get nervous. At that point, we try our best to calm them down by giving positive feedback, helping them relax and feel more confident''. P23 reinforced this sentiment, affirming: ``I believe that during the presentations is when people showed their feelings the most, and we tried to lighten the mood or add to the discussion in some way so that everyone felt more comfortable''.

\subsubsection{Managing Emotions (the ability to regulate emotions to promote emotional and intellectual growth)}
An issue among participants was the \textbf{prior planning to mitigate stress}. They found that structured approaches helped them stay focused and emotionally balanced. P3 highlighted, ``The most challenging part was the deadlines, which were quite complicated at first because the deadlines and demands were close to my university assignments, requiring me to be resilient in meeting those deadlines. My strategy was to create a schedule (in Google Calendar) for studying and managing my submissions''. In turn, P24 added, ``I always follow a 'Natural Pomodoro'\footnote{A time management strategy that divides work into focused intervals.} style, taking breaks to rest and clear my mind''. This learning-oriented approach emphasizes how acknowledging and adapting to emotional challenges can enhance long-term emotional resilience and productivity. Similarly, P13 shared the following reflection: "I make everything clear in a to-do list with the dates and priorities for each task. It’s difficult for me to regulate anxiety, so I use breathing exercises and keep the mindset of just do one thing at a time''. Combining task management with mindful breathing may demonstrate how personal organization can lay the groundwork for effective emotional regulation in students \cite{jarvenoja2019emotional}.

Students also demonstrated an adaptive approach to emotional challenges, discussing how to \textbf{channel positive and negative emotions constructively}. P14 shared, ``My strategy is to envision a better future and use positive emotions as fuel; however, if the emotions are negative, I aim to learn from the experience''. This dual strategy of leveraging positive emotions for motivation while extracting lessons from negative ones highlights an adaptive approach. P6 reflected, ``During the end-of-semester period, my emotions were a mess, but I tried to organize myself so that it wouldn't significantly affect my performance. This organization was through schedules that outlined the timing and importance of each task''.

The importance of \textbf{seeking support and fostering collaboration} was another finding. P7 mentioned, ``Regarding the deadlines, a simple individual agenda wasn't helping, so we created a collaborative task schedule with deadlines on Trello, specifically for the team members, outlining what had been completed and what was still in progress''. This strategy of using collaborative tools facilitated task management and provided a support network to help alleviate individual stress. P20 reverberated: ``Always trying to stay calm and remember that we're all there to learn together. By understanding this, we aimed to give our best in what we were assigned to do''. P18 further emphasized: ``I used these emotions to analyze my main weaknesses and to learn how to cope with these feelings''. By emphasizing the collective learning process and the value of teamwork, students demonstrated how a supportive environment can mitigate personal stress.

Finally, students highlighted the relevance of \textbf{balancing between work demands and personal well-being}. P21 shared, ``I take breaks and engage in activities I enjoy that are unrelated to work. Listening to music while tackling tasks also helps me relax and concentrate''. Hence, integrating relaxation techniques and hobbies into the work routine underscores the importance of self-care in managing emotional responses. P4 also added, ``I relied a lot on my nervousness and anxiety as a driving force to overcome the difficulties. In a way, I'm still in the process of self-discovery regarding my limits at the end of the semester, which is why I think I didn't utilize the best strategies''. This finding reflects the ongoing challenge of balancing intense work demands with emotional regulation.

\section{Discussion}

In general, the findings of this work emphasize the \textbf{impact that EI) have on stress management, creativity, and teamwork} within a SE educational setting based on the use of project-based learning. In this sense, students highlighted the importance of structured planning as a means to mitigate stress, with many employing tools like Google Calendar and to-do lists to break down complex tasks into manageable work. This approach facilitated task completion and served as a coping mechanism, reducing anxiety and fostering a sense of control over the workload. These strategies align with the broader literature on stress management \cite{glazer2017work}.%

In addition, our study revealed an understanding of \textbf{how students leveraged both positive and negative emotions to enhance their work processes}. Positive emotions were frequently used as motivational fuel, propelling students through challenging tasks and fostering cognitive flexibility. The ability to provide and receive feedback during high-pressure situations, such as presentations, also emerged as an important facet of EI. Students emphasized how positive reinforcement helped alleviate stress and promoted a collaborative learning environment. This finding aligns with the broaden-and-build theory \cite{fredrickson2004broaden}, which posits that positive emotions expand individuals' thought-action repertoires, leading to greater creativity and problem-solving capabilities. The findings, furthermore, strengthen Graziotin et al.'s ~\cite{graziotin2015How} theory of affect and its impact on programming performance, where positive emotions are described as catalysts on performance in daily programmer activities. On the other hand, negative emotions, rather than being wholly detrimental, were often repurposed as catalysts for perseverance and commitment in the perspective of the students. Rather than viewing emotions as obstacles, students approached them as opportunities for growth. Techniques like the `Pomodoro' method, which integrates regular breaks into work sessions, were used to maintain focus \cite{ren2021steadi}. This adaptive mindset highlights the potential for developing emotional resilience over time in the learning journey \cite{hammond2004impacts, tarbetsky2017social, cahill2020social}.

This study also uncovered \textbf{situations where EI was less effective or where students struggled to apply it}. For example, students reported difficulties in accurately interpreting emotional cues from peers they were less familiar with, which sometimes led to misunderstandings or conflicts. There were instances where stress or nervousness overwhelmed some participants, leading to emotional disengagement and decreased motivation. These situations highlight areas where additional support or targeted training in EI could enhance its application in cooperative learning environments \cite{chapin2015effect}. Addressing these challenges aligns with prior findings that suggest emotional self-regulation and conflict resolution strategies as critical yet underdeveloped skills in educational settings \cite{valente2020conflict}.

We also found that students acknowledged the \textbf{importance of perceiving and interpreting colleagues' emotions}, particularly during stressful periods. In other words, understanding the emotional states of peers allowed participants to provide timely support, thereby fostering a collaborative and empathetic environment. As clarified by Lehner \cite{lehner2020teamwork}, EI is a fundamental factor in enhancing teamwork and project outcomes. Moreover, the ability to empathize and respond to the emotional needs of others not only strengthened relationships but also improved the overall efficiency and effectiveness of team tasks. We also discussed the \textbf{role of self-care practices} \cite{smith2017self}, such as taking breaks and engaging in enjoyable activities, in maintaining emotional balance and productivity. Students who integrated relaxation techniques (e.g, listening to music) reported better  ability to manage stress \cite{sharma2014mindfulness, lata2021listening}.

Our last analysis explored \textbf{the core EI abilities proposed by Mayer and Salovey} \cite{mayer1997emotional}. Concerning the ability to \textit{perceive emotions} in oneself and others was evident in students' reflections on recognizing and interpreting both their own and their peers' emotional states. This ability facilitated more effective teamwork and personal stress management. In addition, the \textit{use of emotions} to aid in cognitive processes, such as problem-solving and creativity, demonstrated the emotional facilitation of thought. Students utilized positive emotions to maintain motivation, highlighting the how emotional states could be used to enhance mental processes. Furthermore, the \textit{understanding about emotions} was reflected in how students understood complex emotional issues within their teams. In particular, by interpreting emotional cues and approaching them to guide interactions and decisions alongside the project, students displayed a certain degree of comprehension of how emotions influenced relationships and work dynamics.  Lastly, the ability to \textit{manage emotions}, both personally and within the team, played a key role in maintaining productivity and fostering a supportive environment during the intense pace of the project-based experience. Strategies such as relaxation techniques and seeking support highlighted the importance of moderating emotional responses to sustain engagement.

\begin{mdframed}[backgroundcolor=mygray]
\textbf{Response to RQ: \textit{What are the self-perceptions of Software Engineering students regarding their Emotional Intelligence after participating in a cooperative learning experience}?} Students generally found it easier to identify their own emotions than to interpret the emotions of others, attributing this ability to self-awareness and prior experience. They also recognized the positive impact of emotions on creativity, problem-solving, collaboration, and communication within their squads. Students acknowledged challenges in interpreting the emotions of others, especially those they were less familiar with, and emphasized the importance of empathy and emotional awareness in maintaining motivation and learning. 
\end{mdframed}

\subsection{Implications for SE Researchers and Educators}

This study shares findings about the understanding of EI in cooperative learning environments. Previous work has explored various aspects, including integrating EI into SE curricula \cite{hidalgo2004use, arroyo2019development}, the intersection between EI and work performance \cite{kosti2014personality, Khatun2020moderation, wrobel2013emotions}, the influence of emotional awareness on productivity \cite{fountaine2017emotional, girardi2020recognizing, girardi2021emotions, graziotin2018happens}, and the impact of EI on software engineering practices and processes \cite{luong2021agile, madampe2024supporting}. Our work extends this knowledge by focusing on EI as a multidimensional construct in the viewpoint of the own SE students within a project-based learning setting, aligning with the model proposed by Mayer and Salovey \cite{mayer1997emotional}.

For SE educators, this study emphasizes the importance of examining the intersection of EI with factors that affect educational outcomes such as team composition, gender inclusiveness, well-being, and analogous real-world project complexities. Practical recommendations include incorporating activities like structured reflections on emotional experiences, role-playing to build empathy, and workshops on self-regulation and conflict management. Encouraging the use of planning tools and stress-management strategies identified can help students develop coping mechanisms. Integrating EI assessments into team evaluations further highlights the importance of emotional competencies in collaboration. This approach could lead us towards tailored EI training programs that enhance emotional skills, improve performance, and better prepare students for the emotional demands of professional SE.

\subsection{Key Takeaways for Industry Practitioners}

For industry practitioners, this study serves as evidence of the relevance of EI in cultivating effective teamwork and managing stress, particularly among novice developers. Key takeaways include the need for structured planning and emotional awareness to sustain productivity and reduce anxiety \cite{navas2018emotional}. Companies would consider integrating EI training into professional development programs to enhance employees' emotional management skills, leading to improved team dynamics, creativity, and problem-solving \cite{frias2021impact, ivcevic2007emotional}.

Our research also discusses the value of creating a supportive environment that acknowledges and addresses emotional challenges \cite{sujila2023effect}. By leveraging EI in the workplace, organizations may enhance individual well-being and overall effectiveness, leading to improved software products and project outcomes. As noted by Novielli and Serebrenik \cite{novielli2019sentiment}, recent SE research has already shown that positive affect correlates with increased productivity. For example, Girardi et al. \cite{girardi2021emotions} provided empirical evidence linking emotions with perceived productivity in the workplace. Furthermore, Graziotin et al. \cite{graziotin2018happens} examined both the positive and negative effects of happiness and unhappiness on developers' mental well-being and the software development processes/artifacts. %

\section{Threats to Validity}

\textbf{\textit{Internal Validity}.} Internal validity threats include reliance on self-reported data, which may lead to self-selection bias, as motivated students with heightened self-awareness might be more likely to participate. Assigning participants to squads based on shared interests could influence their motivation. The mix of high school and undergraduate students may have impacted their experiences. Recall and social desirability biases are also concerns, as participants might report favorable rather than genuine responses. To enhance credibility, we employed data triangulation through questionnaires, focus groups, and literature. We ensured reflexivity through collaborative discussions of assumptions, biases, and findings.

\textbf{\textit{External Validity.}} Our study focused on a specific group of participants with similar educational background, which inherently limits the generalizability of the findings. This limitation arises from our use of purposeful sampling, which prioritizes depth and richness of information. However, generalizability was not the primary aim of this research; rather, we sought to gain an in-depth understanding of students' self-perceptions. To enhance transferability, we provided detailed descriptions of the context, participants, and method.

\textbf{\textit{Construct Validity}.} The measurement of EI poses a threat to construct validity. Instead of using psychometric tools, we inferred EI from participants' narratives, which may not fully capture its scope as defined in psychological literature. To establish a shared understanding, we began the focus groups with a brief explanation of EI, its importance, and ways to develop it. However, this explanation might have introduced bias, as participants could align responses with perceived researcher expectations. To mitigate this threat, our questionnaire and focus group scripts were grounded in Mayer and Salovey's seminal studies \cite{mayer1997emotional, mayer2001emotional, mayer2008human}.

\textbf{\textit{Conclusion Validity}.} While our study provides findings into SE students' self-perceptions of EI after a cooperative learning experience, it does not aim to establish causal relationships. The interplay of factors such as team dynamics, individual differences, and project-specific challenges may have influenced the observed effects, making it difficult to determine their significance. However, this context aligns with our objective of exploring students' self-perceptions of EI rather than identifying direct causal links.

\section{Conclusions and Future Work}

This study examined Software Engineering (SE) students' self-perceptions of Emotional Intelligence (EI) following a two-month cooperative learning experience, focusing on core EI abilities identified by Mayer and Salovey \cite{mayer1997emotional}. Using questionnaires and focus groups, we assessed students' familiarity with EI, their self-evaluation of emotional skills, and their views on EI’s impact on academic performance and teamwork. Findings showed how students coped with stress, pressure, and conflicts, including their perception, use, understanding, and management of emotions throughout the course.

This research contributes to both the academic and practical understanding of EI in SE. Academically, it extends the application of EI theories, particularly Mayer and Salovey's model, by providing empirical evidence of its relevance in the SE educational context. Practically, we offer key takeaways for SE teams and educators, emphasizing the importance of cultivating EI abilities to enhance project outcomes. By highlighting the positive impact of EI on creativity, stress management, and collaboration, this research advocates for developing EI capabilities as part of SE educational curricula

Future work could focus on longitudinal studies tracking EI development in SE students, explore how EI influences the handling of real-world project challenges and stressors, and investigate the impact of gender on EI, including issues like toxic culture and impostor syndrome. Additionally, further research could assess strategies for integrating EI into modern SE curricula.

{\footnotesize\bibliography{references}}
\bibliographystyle{IEEEtran}

\end{document}